\documentstyle[twoside,fleqn,npb,epsfig]{article}

\title{Perturbative and Non-perturbative Lattice Calculations 
       for the Study of Parton Distributions}

\author{Stefano Capitani\address{DESY Zeuthen, 
        John von Neumann-Institut f\"ur Computing (NIC)\\
        Platanenallee 6, 15738 Zeuthen, Germany}
        \thanks{Presented at RADCOR 2002 and Loops and Legs 2002, 
                Kloster Banz (Germany), September 8 to 13, 2002.}}

\begin{document}

\begin{abstract}
We discuss how lattice calculations can be a useful tool for the study 
of structure functions. Particular emphasis is given to the perturbative
renormalization of the operators.
\end{abstract}

\maketitle

\section{INTRODUCTION}

Lattice QCD provides non-perturbative techniques for the computation of the 
Mellin moments of the structure functions of hadrons from first principles, 
without model assumptions. Many results have been obtained in the past years 
from the lattice, and include the calculation of the lowest moments of various
structure functions of the quarks: the unpolarized structure functions, 
the spin-dependent structure functions $g_1$ and $g_2$, and the transversity 
structure function $h_1$.

Lattice perturbation theory is essential for the renormalization of the 
relevant operators. Perturbative calculations of renormalization factors
have been carried out for the lowest three moments of all structure 
functions, in the case of Wilson and also of overlap fermions (which do not 
break chiral symmetry). For many operators these renormalization factors 
are also known in the improved theory.
Perturbative renormalization factors have also been calculated 
for some higher-twist matrix elements (4-quark operators).

\section{STRUCTURE FUNCTIONS} 

It is not possible to compute a complete structure function directly 
on the lattice (which is set up in Euclidean space).
The reason is that the structure functions describe the physics 
close to the light cone, and this region of Minkowski space shrinks to
a point when one goes to Euclidean space, where Monte Carlo simulations
are performed. However, on a Euclidean lattice it is possible to compute
the moments of the structure functions, using the operator product expansion: 
\begin{displaymath}
\int_0^1 x^n {\cal F}_{(i)} (x,Q^2) \sim C^{(n,i)} (\frac{Q^2}{\mu^2})
\cdot \langle h | O^{(n,i)} (\mu)  | h \rangle .
\end{displaymath}
The Wilson coefficients contain the short-distance physics, and can be
perturbatively computed in the continuum. The matrix elements contain the
long-distance physics, and can computed using numerical simulations, 
supplemented by a lattice renormalization of the relevant operators.

Operators of twist two (twist: dimension minus spin) 
dominate the expansion above. Moments of the unpolarized 
structure functions (which give the unpolarized distribution $q$), 
of the spin-dependent structure functions $g_1$ (which gives the 
helicity distribution $\Delta q$) and $g_2$, and of the transversity 
structure function $h_1$ (which gives the transversity distribution $\delta q$)
are measured by towers of hadronic matrix elements:
\begin{eqnarray*}
\langle x^n \rangle &\sim& \langle  h  | 
\, \overline{\psi}  \gamma_{\{\mu} D_{\mu_1} \cdots D_{\mu_n\}}  \psi
\,  |  h  \rangle \\
\langle (\Delta x)^n \rangle &\sim& \langle  h  |  
\, \overline{\psi}  \gamma_5 \gamma_{\{\mu} D_{\mu_1} \cdots D_{\mu_n\}}  \psi
\,  |  h  \rangle \\
\langle x^n \rangle_{g_2} &\sim& \langle  h  |  
\, \overline{\psi}  \gamma_5 \gamma_{[\mu} D_{\{\mu_1} D_{\mu_2]}
\cdots D_{\mu_n\}}  \psi \, 
 |  h  \rangle \\
\langle (\delta x)^n \rangle &\sim& \langle  h  | 
\, \overline{\psi}  \gamma_5 \sigma_{\mu\{\mu_1} D_{\mu_2} 
\cdots D_{\mu_n\}}  \psi \, 
 |  h  \rangle .
\end{eqnarray*}

\section{OPERATOR RENORMALIZATION}

To obtain physical continuum matrix elements from the lattice, 
the results of Monte Carlo simulations need to be renormalized 
from the lattice to a continuum scheme. 
To obtain these renormalization factors one has to compute 
1-loop matrix elements on the lattice as well as in the continuum.
The lattice operators at tree level, for $p \ll \pi/a$, have the same 
matrix elements as the original continuum operators, so that at 1 loop
\begin{eqnarray*}
\langle O_i^{\rm lat} \rangle &=& 
\frac{g_0^2}{16 \pi^2} \Big( -\gamma_{ij} \log a^2p^2 + R_{ij}^{\rm lat} 
\Big) \langle O_j^{\rm tree} \rangle \\
\langle O_i^{\rm cont} \rangle &=& 
\frac{g_{\overline{\rm{MS}}}^2}{16 \pi^2} \Big( -\gamma_{ij} 
\log \frac{p^2}{\mu^2} 
+ R_{ij}^{\rm cont} \Big) 
\langle O_j^{\rm tree} \rangle ,
\end{eqnarray*}
where for the continuum one chooses the $\overline{\rm{MS}}$ scheme,
since the Wilson coefficients are known in this scheme.
In general $R_{ij}^{\rm lat} \neq R_{ij}^{\rm cont}$, and the 1-loop 
renormalization factors on the lattice and in the continuum are different.
The anomalous dimensions are equal, and thus for $\mu=1/a$
only a finite renormalization connects the two schemes.
The matching between Monte Carlo numbers and the 
renormalized physical results is then 
\begin{displaymath}
\langle O_i^{\rm cont} \rangle = \Big( \delta_{ij} 
-\frac{g_0^2}{16 \pi^2} \Big( R_{ij}^{\rm lat} -R_{ij}^{\rm cont} \Big) \Big) 
\langle O_j^{\rm lat} \rangle .
\end{displaymath}

Since Lorentz symmetry is broken on the lattice (which is invariant under 
the hypercubic group), and for Wilson fermions also chiral symmetry, 
the mixing of operators under renormalization is more complicated than 
in the continuum theory. Operators multiplicatively renormalizable in 
the continuum may mix with other operators when they are put on the lattice. 
Sometimes these mixing coefficients are even power divergent, that is they 
behave like $1/a^n$. In general then the matching factors 
$R_{ij}^{\rm lat} -R_{ij}^{\rm cont}$ are not square matrices.

\section{WILSON FERMIONS}

Most calculations are done using the discretization of the (Euclidean) 
QCD action proposed by Wilson. Wilson fermions however break chiral symmetry, 
and this causes an additive mass renormalization even for a zero 
bare mass, and a heavy pion ($m_\pi \sim 500~MeV$).
Extrapolations to the chiral limit are then needed, and this  
source of systematic errors has to be controlled.

Lattice perturbation theory is quite cumbersome: there are more vertices 
and more diagrams than in the continuum, the expressions contain a huge 
number of terms, and the integrals are more complicated.
The Wilson quark propagator is
\begin{displaymath}
\delta^{ab} a 
\frac{- {\rm i} \sum_\mu \gamma_\mu \sin a k_\mu + a m_f + 2 r 
\sum_\mu \sin^2 {\displaystyle \frac{a k_\mu}{2}} }{ \sum_\mu \sin^2 a k_\mu +
\Big[ 2 r \sum_\mu \sin^2 {\displaystyle \frac{a k_\mu}{2}} + a m_f \Big]^2} ,
\end{displaymath}
which in the continuum limit gives
\begin{displaymath}
\delta^{ab} \, \frac{-{\rm i} \sum_\mu \gamma_\mu k_\mu + m_f}{\sum_\mu k_\mu^2
+ m_f^2} ,
\end{displaymath}
and the gluon propagator (in covariant gauge) is
\begin{displaymath}
\frac{\delta^{ab}}{{\displaystyle \frac{4}{a^2}} 
\sum_\lambda \sin^2 {\displaystyle \frac{a k_\lambda}{2}} } 
\Bigg[\delta_{\mu\nu} 
-(1-\alpha)  \frac{ \sin {\displaystyle \frac{a k_\mu}{2}}  
\sin {\displaystyle \frac{a k_\nu}{2}}    }{
\sum_\lambda \sin^2 {\displaystyle \frac{a k_\lambda}{2}} } \Bigg] .
\end{displaymath}

The lattice theory has exact gauge invariance at any finite $a$.
This causes the presence in the action of the group elements 
$U_\mu = \exp ({\rm i}ag_0 A_\mu)$ instead of the algebra elements $A_\mu $. 
One has then to expand the $U_\mu$'s in terms of the $A_\mu$'s, and this 
generates an infinite number of vertices, of which only a finite number is 
needed at any given order in $g_0$. The quark-quark-gluon vertex is
\begin{equation}
- g_0 (t^a)^{bc}
\Big( {\rm i} \gamma_\mu \cos \frac{as_\mu}{2}
+ r \sin \frac{as_\mu}{2} \Big)
\label{eq:qqg}
\end{equation}
(where $s=p_1+p_2$, the momenta of the quarks),
which for $a \to 0$ gives $- g_0 (t^a)^{bc}  {\rm i} \gamma_\mu$, 
while the quark-quark-gluon-gluon vertex is
\begin{equation}
- \frac{1}{2} a g_0^2  \delta_{\mu\nu} \{ t^a, t^b \}^{cd} 
\Big( -{\rm i} \gamma_\mu \sin \frac{as_\mu}{2}
+ r \cos \frac{as_\mu}{2} \Big) \label{eq:qqgg}.
\end{equation}
This is an irrelevant vertex (is zero in the continuum limit), but still gives 
non-vanishing contributions to Feynman diagrams in divergent loops.

Also in the pure gauge part there is an infinite number of vertices.
The 3-gluon vertex is 
\begin{eqnarray*}
\lefteqn{ {\rm i} g_0  f^{abc}  \frac{2}{a} \Bigg\{
    \delta_{\mu\nu}  \sin \frac{a (k-p)_\rho}{2} \cos \frac{a q_\mu}{2} } \\
&& +\delta_{\nu\rho} \sin \frac{a (p-q)_\mu}{2}  \cos \frac{a k_\nu}{2} \\
&& +\delta_{\rho\mu} \sin \frac{a (q-k)_\nu}{2}  \cos \frac{a p_\rho}{2} 
\Bigg\} ,
\end{eqnarray*}
which in the continuum limit becomes
\begin{displaymath}
{\rm i} g_0  f^{abc}  \Big\{
    \delta_{\mu\nu}  (k-p)_\rho 
   +\delta_{\nu\rho}  (p-q)_\mu
   +\delta_{\rho\mu}  (q-k)_\nu \Big\}.
\end{displaymath}
The 4-gluon vertex is too complicated to be reported here. Furthermore, 
the gauge measure at order $g_0^2$ gives a $1/a^2$ mass counterterm.
Finally, using a Faddeev-Popov procedure one can obtain the Feynman rules for
the ghost propagator and the ghost interactions.
The effective ghost-gauge field interaction is not linear in the gauge 
potential $A_\mu$, and thus also in this sector new vertices appear that have 
no continuum analog, like the ghost-ghost-gluon-gluon vertex.

\section{OVERLAP FERMIONS}

A Dirac operator $D = \gamma_\mu D_\mu$ which satisfies the Ginsparg-Wilson 
relation $\gamma_5 D + D \gamma_5 = a/\rho D \gamma_5 D$ defines fermions 
with maintain an exact chiral symmetry also for non-zero lattice spacing 
and also keep all other fundamental properties.

One of the possible solutions of the Ginsparg-Wilson relation is given by
overlap fermions. In the massless case the overlap-Dirac operator is
\begin{displaymath}
D_N = \frac{1}{a} \rho  \Big[ 1+ \frac{X}{\sqrt{X^\dagger X}} \Big] ,
\qquad X=D_W -\frac{1}{a} \rho ,
\end{displaymath}
where $D_W$ is the Wilson-Dirac operator.
For $0 <\rho <2r$ one has the right spectrum of massless fermions.
Additive mass renormalization is then forbidden, and one avoids 
a source of systematic errors always present using Wilson fermions.

The massless quark propagator in the overlap is more complicated than 
in Wilson, and is
\begin{displaymath}
\delta^{ab}  \Bigg(
\frac{-{\rm i} \sum_\mu \gamma_\mu \sin ak_\mu}{2 \rho
\left( \omega(k)+b(k) \right) } + \frac{a}{2 \rho} \Bigg) ,
\end{displaymath}
where $b(k) = 1/a \, ( 2r \sum_\mu \sin^2 ak_\mu/2 -\rho )$, \newline
$\omega(k) = \sqrt{X_0^\dagger (k) X_0 (k)}$,
and 
\begin{displaymath}
X_0 (k)=\frac{1}{a} \Big( {\rm i} \sum_\mu \gamma_\mu \sin ak_\mu
+ 2r \sum_\mu \sin^2 \frac{ak_\mu}{2} -\rho \Big)  .
\end{displaymath}

The overlap vertices can be expressed in terms of the vertices of the 
QED Wilson action $W_{1\mu}$ and $W_{2\mu}$, which are the vertices 
(\ref{eq:qqg}) and (\ref{eq:qqgg}) without the color matrices, and of the 
quantity $X_0$. For example, the quark-quark-gluon vertex is
\begin{eqnarray*}
\lefteqn{ (t^a)^{bc}  \cdot 
\rho  \frac{1}{\omega(p_1) + \omega(p_2)} 
\Bigg[ W_{1\mu} (p_1,p_2)}  \\
&& \qquad -\frac{1}{\omega(p_1)\omega(p_2)} X_0(p_2) 
W_{1\mu}^\dagger (p_1,p_2) X_0(p_1) \Bigg] .
\end{eqnarray*}
The quark-quark-gluon-gluon vertex is very long and will not be given here
(see for example~\cite{pertrenorm_ov}).

\section{IMPROVEMENT}

It is very expensive to decrease the errors due to the granularity of 
the lattice just by reducing the lattice spacing $a$, because the cost 
would grow like $a^{-5}$ in the quenched theory, and in full QCD it 
would grow even faster. Halving the discretization errors in this way 
would then be at least $30$ times more expensive.
A more effective way is to improve actions and operators.

$O(a)$ improvement removes the contributions of order $a$ to the systematic 
error arising from the finiteness of the lattice spacing
by adding a counterterm to the fermion action, so that 
\begin{displaymath}
\left< p \left| \widehat{\cal O}_{L} \right| p' \right>_{MC}=
a^{d} \left[ \left< p \left| \widehat{\cal O} \right| p' 
\right>_{ph} + O(a^2) \right] .
\end{displaymath} 
For Wilson fermions, this is achieved in on-shell matrix elements adding 
a counterterm whose coefficient $c_{sw}$ has to be exactly tuned. 
Only for its appropriate value, for a given $g_0$, the $O(a)$ 
effects are canceled and one gets a faster convergence to the continuum limit.
Due to this action counterterm one has to add to the Wilson 
vertex (\ref{eq:qqg}) the improved quark-quark-gluon interaction 
vertex~\footnote{The fermion propagator and the vertices with an even number 
of gluons are instead not modified by the improvement.
Neither is the gluon propagator: the first corrections to the pure gauge 
action are already of order $a^2$.}
\begin{displaymath}
- c_{sw} \cdot g_0 \frac{r}{2} (t^a)^{bc} 
\cos \frac{aq_\mu}{2} \sum_\lambda
\sigma_{\mu\lambda} \sin aq_\lambda ,
\end{displaymath}
where $q$ is the difference between the incoming and outgoing momenta
of the quarks.
In addition to this, one also has to improve the various operators,
that is one must add a basis of higher dimensional (irrelevant) operators 
with the same quantum numbers, with $\dim (\widetilde{O}_i) = \dim (O) + 1$:
\begin{displaymath}
O  \longrightarrow  O^{\rm imp} 
= (1+b_O  a m)  O + a \sum_i c_i  \widetilde{O}_i .
\end{displaymath}

The $O(a)$ improvement for overlap fermions is much simpler. 
The action is already improved, and thus there is no need of new interactions.
The improved operators are given, to all orders, by 
\begin{displaymath}
O^{\rm imp} = \bar{\psi} \Big(1-\frac{1}{2\rho} a D_N\Big)  
\widetilde{O}  \Big(1-\frac{1}{2\rho} a D_N\Big) \psi 
\end{displaymath}
(while for Wilson fermions the improvement construction is different for 
each operator). One thus gets full $O(a)$ improvement without tuning any 
coefficients, and to all orders of perturbation theory.
For Wilson fermions, instead, one has to determine the coefficients of 
the operator counterterms order by order in perturbation 
theory.~\footnote{This appears to be a difficult task, and even for 
the first moment of unpolarized distributions all lowest-order improvement 
coefficients have not yet been determined.}

\section{FORM CODES}

Due to the complexity of the calculations, to the possible great number of
diagrams, and above all to the huge amount of terms in each diagram, 
computer codes have to be used.
To evaluate the Feynman diagrams and obtain the algebraic expressions 
for the renormalization factors, we have thus developed sets of computer codes 
written in the symbolic manipulation language FORM.
These codes take as input the Feynman rules for the particular 
combination of operators, propagators (Wilson or overlap) and vertices 
(Wilson, improved, or overlap) appearing in each diagram, expand 
them in the lattice spacing $a$ at the appropriate order, evaluate the 
gamma algebra on the lattice, and then work out everything until the final 
expressions are obtained.
The main difficulty in developing such codes is that the (Euclidean) Lorentz 
group $O(4)$ breaks down to the hypercubic group $W_4$.
A representation which is irreducible under the (Euclidean) Lorentz group
is thus in general reducible under the symmetry group of the lattice.
This gives rise to a whole new class of problems, of which the most serious
concerns the Einstein summation convention.

The FORM language has been developed having in mind the usual continuum 
calculations. There are therefore many useful built-in features which are 
sometimes somewhat of an hindrance when one tries to perform lattice 
perturbative calculations. A blind use of them would give, for example,
\begin{displaymath}
\sum_{\lambda} \gamma_{\lambda} p_{\lambda} \sin k_{\lambda} 
\longrightarrow p \! \! / \sin k_{\lambda} .
\end{displaymath}
One thus needs to develop special routines to cope with the gamma algebra on 
the lattice.

Integrals are done over the first Brillouin zone
\begin{displaymath}
\int^{\frac{\pi}{a}}_{-\frac{\pi}{a}} \frac{d^4 k}{(2\pi)^4} \, f(ak, ap) 
\, \rightarrow \, \int^\pi_{-\pi} \frac{d^4 k'}{(2\pi)^4} \, a^{-4} f(k', ap)
\end{displaymath}
with the rescaling $k' = a k$, and in general an expansion in $ap$ is needed. 
FORM codes become then necessary also because of the huge number of terms 
arising from the Taylor expansions of the Fourier transforms of operators, 
propagators, vertices and covariant derivatives. 
The $n$-th moment of a parton distribution behaves like
\begin{displaymath}
\langle \overline{\psi}  \Gamma
 D_{\mu_1} \cdots  D_{\mu_n}  \psi \rangle ~~~\sim~~~\frac{1}{a^n} ,
\end{displaymath}
because $D \sim 1/a$, and thus one has to perform an expansion in $a$ to order 
$n$ for every quantity. For example, the Wilson quark propagator to $O(a)$ is
\begin{eqnarray*}
\lefteqn{ \delta^{ab} D^{-1}(k) \Bigg\{
\Big( - {\rm i} \sum_\mu \gamma_\mu \sin k_\mu + 2 r 
\sum_\mu \sin^2 {\displaystyle \frac{k_\mu}{2}} \Big) } \\
\lefteqn{ + a \cdot \Bigg[ 
\Big( - {\rm i} \sum_\mu \gamma_\mu q_\mu \cos k_\mu 
+ r \sum_\mu q_\mu \sin k_\mu + m_f \Big) } \\
&& - D^{-1}(k) \Big( - {\rm i} \sum_\rho \gamma_\rho \sin k_\rho 
+ 2 r \sum_\rho \sin^2 {\displaystyle \frac{k_\rho}{2}} 
\Big) \\
&& \times \Big( 2 \sum_\mu q_\mu \sin k_\mu \cos k_\mu \\
&& \, \,  + 4 r \sum_\mu \sin^2 {\displaystyle \frac{k_\mu}{2}} 
\Big( r \sum_\nu q_\nu \sin k_\nu + m_f \Big) \Big)
\Bigg] \Bigg\} ,
\end{eqnarray*}
where the denominator is
\begin{displaymath}
D(k) = \sum_\mu \sin^2 k_\mu +
\Big( 2 r \sum_\mu \sin^2 {\displaystyle \frac{k_\mu}{2}} \Big)^2 .
\end{displaymath}
As a consequence, a huge number of terms appear (at least in the initial 
stages of the manipulations), and also the gamma algebra becomes quite 
cumbersome to do by hand. All this also implies a limitation on the number 
of moments of structure functions that one can compute.~\footnote{This comes 
on top of the limitation coming from operator mixing: when an operator 
has more than four indices, two of them have to be equal, and this opens 
the door to lots of mixings, often with power-divergent coefficients.}

\section{RESULTS}

We have calculated the renormalization in the $\overline{\rm{MS}}$ scheme 
of several operators measuring the lowest moments of structure 
functions~\cite{pertrenorm,pertrenorm_ov}.
In some cases for a given moment we have computed two operators,
labeled $(a)$ and $(b)$, which belong to two different representations of the 
discrete Euclidean Lorentz group (see Table~\ref{fig:renorm}).~\footnote{In 
the operators $(a)$ all Lorentz indices are distinct.}
Operator mixing for overlap fermions is simpler than for Wilson fermions:
chiral symmetry is not broken and thus it prohibits any mixing with operators 
of different chirality. Mixing coefficients which are power-divergent like 
$a^{-n}$ in the continuum limit can be eliminated from the start from overlap 
calculations if the corresponding operators belong to multiplets with the 
wrong chirality. In the Wilson case, indeed, the operators measuring 
the moments of the $g_2$ structure function present additional mixings with 
wrong-chirality operators with power-divergent coefficients. Furthermore, 
the $Z$'s of the $n$-th moments of $q$ and $\Delta q$ are not constrained 
to be equal anymore.

\begin{table}[t]
\begin{center}
  \begin{tabular}{crrr} \hline
moment   &  \multicolumn{2}{c}{overlap} & Wilson \\ 
  \cline{2-3}
         &  \multicolumn{1}{r}{$\rho=1.0$} &  
            \multicolumn{1}{r}{$\rho=1.9$} &   \\ \hline
$\langle x \rangle_q^{(a)} $ &
1.41213  &  1.21841  &  0.98920  \\
$\langle x \rangle_{\Delta q}^{(a)} $ &
1.41213  &  1.21841  &  0.99709  \\
$\langle x \rangle_q^{(b)} $ &
1.40847  &  1.21309  &  0.97837  \\
$\langle x \rangle_{\Delta q}^{(b)} $ &
1.40847  &  1.21309  &  0.99859  \\
$\langle x^2 \rangle_q $ &
1.51968  &  1.32436  &  1.09763  \\
$\langle x^2 \rangle_{\Delta q} $ &
1.51968  &  1.32436  &  1.10231  \\
$\langle x^3 \rangle_q^{(a)} $ &
1.61872  &  1.42279  &  1.19722  \\
$\langle x^3 \rangle_{\Delta q}^{(a)} $ &
1.61872  &  1.42279  &  1.20040  \\
$\langle x^3 \rangle_q^{(b)} $ &
1.63737  &  1.44159  &  1.21534  \\
$\langle x^3 \rangle_{\Delta q}^{(b)} $ &
1.63737  &  1.44159  &  1.21944  \\
$\langle x \rangle_{g_2} $ &
1.34794  &  1.18456  &  mixing   \\
$\langle x^2 \rangle_{g_2} $ &
1.47816  &  1.30997  &  mixing   \\
$\langle x^3 \rangle_{g_2} $ &
1.58943  &  1.41900  &  mixing   \\
$\langle 1 \rangle_{\delta q} $ &
1.27252  &  1.08648  &  0.85631  \\
$\langle x \rangle_{\delta q} $ &
1.41153  &  1.21851  &  0.99559  \\
$\langle x^2 \rangle_{\delta q} $ &
1.51865  &  1.32355  &  1.10021  \\
\hline
\end{tabular} 
\end{center}
\caption{Renormalization constants of multiplicatively renormalized structure 
function operators, for $a = 0.1~fm$, in the $\overline{\rm{MS}}$ scheme.}
\label{fig:renorm}
\end{table}

For reasons of computing power, structure functions have been studied mostly 
in the quenched approximation, and only recently computations in full QCD 
have become feasible. On the lattice the Grassmann variables have to be 
analytically integrated, and the partition function
\begin{displaymath}
Z = \int {\cal D} U {\cal D} \overline{\psi} {\cal D} \psi  
{\rm e}^{-S_g[U] - \overline{\psi} (\not{\hspace{-0.02cm}D}[U]+m_f) \psi}
\end{displaymath}
becomes in the simulations
\begin{displaymath}
Z = \int {\cal D} U   \det (\not{\hspace{-0.12cm}D}[U]+m_f) 
  {\rm e}^{-S_g[U]} ,
\end{displaymath}
which is equivalent to use an effective action
\begin{displaymath}
S_{\rm eff}[U] = S_g[U] - \ln  \det (\not{\hspace{-0.12cm}D}[U]+m_f) .
\end{displaymath}
Quenching amounts to doing simulations with
$\det (\not{\hspace{-0.12cm}D}[U]+m_f) = 1$.
This means that there are no sea quarks: all internal quark loops are 
neglected. Although it looks quite drastic, it is not a bad approximation 
for many physical quantities, and apparently (as it has been understood 
recently) also for structure functions.

Structure functions in full Wilson QCD have been the major progress in the 
past couple of years~\cite{LHPC,QCDSF}. The first studies with dynamical 
Wilson quarks have shown no statistically significant differences between 
quenched and full QCD results. Previously it had been conjectured that 
quenching was the cause of the observed discrepancies of lattice results 
with experiment. Evidently this is not the case, at least for the values 
of Wilson quark masses attainable at present.

The largest discrepancies appear for unpolarized quark distributions.
The spin-dependent $g_1$ and $g_2$ structure functions have 
also been extensively studied, as well as the axial charge 
$g_A = \Delta u - \Delta d$ (which deviates less from experiment) and 
the $h_1$ transversity structure function~\cite{transv}, whose lowest moment
is the tensor charge $\delta u - \delta d$.

One limitation of present simulations is given by the difficulty to compute
numerically diagrams which are disconnected (except through gluon lines). 
The disconnected diagrams are however flavor independent, and do not 
contribute to the difference between $u$ and $d$ structure functions 
(for $SU(2)$ degeneracy).
This means that quantities like $g_A = \Delta u -\Delta d$ and 
$\langle x^n \rangle_{u-d}$ do not receive contributions 
from disconnected diagrams.
Their lattice results seem also to be closer to the experimental numbers.

The chiral extrapolations of the lattice data linear in $m_q \sim m_\pi^2$ 
could also be responsible for the discrepancies with experiment. It has 
been recently suggested that extrapolations using chiral perturbation theory 
could solve these discrepancies (see~\cite{chpt} and references therein). 
In fact, the pion cloud of the nucleon is not adequately described by current 
lattices. Extrapolation formulae which use chiral perturbation theory 
look like 
\begin{eqnarray*}
\lefteqn{\langle x^n \rangle_{u-d} = A_n  \Big[ 1 
- \frac{(3g_A^2+1)}{(4\pi f_\pi )^2}  m_\pi^2  
\ln \Big(\frac{m_\pi^2}{m_\pi^2 + \Lambda^2} \Big) \Big] }\\
&& + B_n  (m_\pi r_0)^2 + C_n  (a/r_0)^2 ,
\end{eqnarray*}
where $\Lambda$ is a phenomenological cutoff related to the size 
of the source generating the pion cloud.

These extrapolation formulae seem able to resolve the discrepancy with 
experiment. However, results can be reproduced only with $\Lambda$'s with a 
range so wide to have no predictive power, and the currently available lattice
data do not even discriminate between linear and chiral perturbation theory 
fits. The problems is that the available pions are not sufficiently light. 
A smaller pion mass ($m_\pi < 250~MeV$) is needed before the parameters of the
chiral expansions can be well determined on the lattice. Such calculations 
require about 8 Teraflops for one year. The next generation of computers, 
coming in a couple of years, will be able to perform these calculations.

There is no doubt that the pion cloud of the proton is very important.
For it to be adequately included in the lattice box and properly measured 
in Monte Carlo simulations, the pion correlation length should be much smaller
than the lattice size. This points to the use of larger lattices.

Higher-twist corrections have also been studied~\cite{ht}. The lattice results
for the twist four ($1/Q^2$) corrections are (in a few particular cases) 
quite small. The renormalization factors of four-quark operators like
$\sum_a   \overline{\psi} \gamma_\mu \gamma_5  t^a \psi \cdot 
\overline{\psi} \gamma_\nu \gamma_5  t^a \psi$
have been calculated in perturbation theory, and the twist-4 results for 
the first moment of the unpolarized pion and proton structure functions 
turn out to be much smaller than the corresponding twist-2 matrix elements 
computed on the lattice, and also smaller than the phenomenological numbers.
These are not however complete and systematic studies, like the ones
concerning leading-twist operators.
Only particular flavor and isospin combinations could be considered, 
to avoid mixing with lower-dimensional operators.

\section{CONCLUSIONS}

The lattice provides invaluable techniques for investigating moments 
of structure functions non-perturbatively from first-principles.
The matching to the $\overline{\rm{MS}}$ scheme is done by computing 
renormalization factors in perturbation theory (Wilson and overlap).
Simulations with Wilson fermions are now also performed in full QCD, 
but many discrepancies between lattice and experiment remain.
More control on the systematic errors is required.

\end{document}